\documentclass[prl,aps,twocolumn,preprintnumbers,amsmath,amssymb,superscriptaddress,showpacs]{revtex4-2}
\usepackage{graphicx}
\usepackage{graphics}
\usepackage{amsmath}
\usepackage{amssymb}
\usepackage{amsfonts}
\usepackage{dsfont}
\usepackage{braket}
\usepackage{color}
\usepackage{braket,slashed}
\usepackage[mathscr]{euscript}
\definecolor{darkblue}{rgb}{0, 0, 0.8}
\usepackage[colorlinks=true, breaklinks=true, linkcolor=red, citecolor=blue, urlcolor=blue]{hyperref} 
\usepackage{hyperref}
\usepackage{natbib}
\usepackage{subfigure}
\usepackage{xfrac}
\usepackage{bm}
\usepackage{kantlipsum}
\usepackage{enumitem}
\usepackage{tikz}
\usepackage{framed}
\usepackage{graphicx}
\usepackage{subfigure}
\usepackage{cleveref}
\usepackage{adjustbox}
\usepackage{xcolor}

\allowdisplaybreaks[1]

\newcommand{\code}[1]{\texttt{#1}}

\newcommand{\e}{\ensuremath{\mathrm{e}}}

\newcommand{\be}{\begin{equation}}
\newcommand{\ee}{\end{equation}}

\newcommand{\bea}{\begin{eqnarray}}
\newcommand{\eea}{\end{eqnarray}}

\usepackage{lineno}
\usepackage{tikz}
\usetikzlibrary{trees,snakes}

\UseRawInputEncoding

\usepackage{bm}
\usepackage{mathtools}
\usepackage{dsfont}
\usepackage{amsfonts}
\usepackage[normalem]{ulem}
\usepackage{url}
\usepackage{wasysym}
\usepackage{bbm}

\begin{document}
 
\title{Resolving Geometric Excitations of Fractional Quantum Hall States}

\author{Yang Liu}
\affiliation{Beijing National Laboratory for Condensed Matter Physics and Institute of Physics,
Chinese Academy of Sciences, Beijing 100190, China.}
\affiliation{School of Physical Sciences, University of Chinese Academy of Sciences, Beijing 100049, China.}

\author{Tongzhou Zhao}
\affiliation{Beijing National Laboratory for Condensed Matter Physics and Institute of Physics,
Chinese Academy of Sciences, Beijing 100190, China.}

\author{T. Xiang}\email{txiang@iphy.ac.cn}
\affiliation{Beijing National Laboratory for Condensed Matter Physics and Institute of Physics, Chinese Academy of Sciences, Beijing 100190, China.}
\affiliation{School of Physical Sciences, University of Chinese Academy of Sciences, Beijing 100049, China.}


\begin{abstract}
The quantum dynamics of the intrinsic metric profoundly influence the neutral excitations in the fractional quantum Hall system, as established by Haldane in 2011 \cite{Haldane2011}, and further evidenced by a recent two-photon experiment \cite{Liang2024}. Despite these advancements, a comprehensive understanding of the dynamic properties of these excitations, especially at long wavelengths, continues to elude interest. In this study, we employ tensor-network methods to investigate the neutral excitations of the Laughlin and Moore-Read states on an infinite cylinder. This investigation deepens our understanding of the excitation spectrum in regions where traditional methods do not work effectively. The spectral functions for both states reveal the presence of $S=-2$ geometric excitations. For the first time, we unveil the complex spectra of both neutral fermion and bosonic  Girvin-MacDonald-Platzman modes within the excitation continuum by calculating the three-particle density response function for the Moore-Read state. Our findings support the hypothesis of emergent supersymmetry and highlight the potential for detecting neutral fermions in future experiments.
\end{abstract}
\maketitle

\textit{Introduction ---}
 Neutral excitations in the fractional quantum Hall effect (FQHE) have drawn intensive attention over the past decades. The pioneering work by Girvin, MacDonald and Platzman \cite{Girvin1985, Girvin1986} interprets the neutral excitation in the Laughlin state \cite{Laughlin1983} as a collective density fluctuation mode, called the magnetoroton, analogous to the roton excitation in superfluid helium \cite{Bijl1941, Feynman1953, Feynman1954, Feynman1956}. The lowest neutral excitation was later identified as a composite fermion exciton mode \cite{Scarola2000, Du1993, Pinczuk1993, Kang2001, Kukushkin2009}.  Haldane proposed a geometric interpretation of this mode in the long-wavelength limit, suggesting that the fluctuation of the intrinsic metric determines its dynamics \cite{Haldane2009, Haldane2011, Haldane2011_1}. More specifically, perturbations on the metric introduce a new excitation in the d-wave channel, possessing both chirality and topological order \cite{Qiu2012, BoYang2012, Yang2016, BoYang2017, Liu2018, Liou2019, Yang2020, Liu2021, Nguyen2021, Nguyen2022, Han2022}, which was observed by circularly polarized resonant inelastic light scattering \cite{Liang2024}.

 The Moore-Read (MR) state \cite{Moore1991, Read1999, Read2000} involves the pairing of composite fermions \cite{Jain1989}, resulting in two types of excitations depending on the electron number: a magnetoroton appears when the electron number is even (even parity), while a neutral fermion emerges when the electron number is odd (odd parity), confirmed by the exact diagonalization~\cite{Moller2011, Bonderson2011, Papi2012}. Unlike the magnetoroton, a neutral fermion carries a half-integer angular momentum \cite{Greiter1991}. The edge states corresponding to the neutral fermions possess non-Abelian statistics. Through quantum interference, they have great potential in the application of topological quantum computing~\cite{Bonderson2009, Bishara2009, Bishara2009_2, Rosenow2009}. 
 
 Recent studies indicate that both types of excitations can be integrated into a cohesive theory if identified as superpartners of an emergent supersymmetry~\cite{Gromov2020} that can be detected from its bulk or edge excitations \cite{Ma2021, Salgado2022}. For the MR state, the two bulk excitations correspond to two edge states known as the chiral charge boson and the copropagating Majorana fermion, associated with $\mathcal{N}=(1,0)$ supersymmetry in $(1+1)$ dimensions \cite{Hull1985, Gates1987}. According to this theory, these excitations should merge in the long-wavelength limit as a manifestation of supersymmetry. However, due to finite size constraints, achieving this limit is challenging with exact diagonalization or other numerical methods.

In this study, we calculate the dynamical spectral functions of the FQHE states on an infinite cylinder using the matrix product state (MPS) renormalization group under the single-mode approximation \cite{Chi2022}. Our results demonstrate that MPS can accurately capture the bosonic magnetoroton and neutral fermion excitations in the continuum, providing more detailed information than the existing theory. Moreover, we find that the neutral magnetoroton mode in the long-wavelength limit is a $S=-2$ geometric excitation (also called a graviton in the literature \cite{Golkar2016a, Gromov2017, Haldane2021, Kirmani2022, Nguyen2023, WangYuZhu2023}). 

{\it Model and method ---} 
 Let us consider a two-dimensional electron gas confined to the lowest Landau level on an infinite cylinder along the x-axis with circumference $L_y$. The parent Hamiltonian of the Laughlin state in the Landau gauge is defined by a two-body interaction 
\be \label{eq:H2}
H_\mathrm{L} = \sum_{k}V_{k}\rho_{k}\rho_{-{k}},
\ee
 where $V_{k}$ is the Fourier transform of a projected pseudopotential \cite{Haldane1983, Trugman1985}. $k= (k_x, k_y)$  is the momentum of electron with $k_y = n e_y$ ($n$ an integer) and $e_y = 2 \pi /L_y$. $k_x$ is continuous on an infinite cylinder. $\rho_k$ is the projected density operator defined in terms of the electron operator $c_n$ in the orbital basis space as 
\be \label{eq:rho}
\rho_{k_x, m e_y } = \sum_n e^{i  \tilde{k}_x (2n + m) / 2 }  c^\dagger_n c_{n+m},
\ee
where $\tilde{k}_x 
= k_x e_y$.

Similarly, the parent Hamiltonian of the MR state is defined by a three-body interaction  \cite{Greiter1991}:
\be \label{eq:H3}
H_\mathrm{MR}= \sum_{{k}_1{k}_2}V_{{k}_1{k}_2} \rho_{{k}_1} \rho_{{k}_2} \rho_{-{k}_1-{k}_2}，
\ee
where $V_{k_1, k_2}$ is the Fourier transform of a three-body interaction \cite{Greiter1991}. Detailed deviations of Eqs.~\eqref{eq:H2} and \eqref{eq:H3} are given in Supplemental Material (SM)~\cite{SM}.

We construct the MPS representation of FQHE states using variational optimization instead of the conventional conformal field theory approach \cite{Estienne2013, Zaletel2012, Crepel2018}. It is noteworthy that the ground states are derived from different orbital configurations \cite{Bernevig2008_1, Bernevig2008_2}. For example, in the 1/3-filling Laughlin state, the ground states are triply degenerate, corresponding to the initial orbital configurations of $\{...100100...\}$, $\{...010010...\}$ and $\{...001001...\}$, respectively. These configurations are translationally invariant with a periodicity of three, effectively acting as a unit cell for these states. Similarly, ground states in other FQHE systems also demonstrate translational invariance, albeit with the same or different unit cell sizes. Thus, the ground state of FQHE can be effectively modeled by a translationally invariant MPS:
\begin{equation} \label{eq:gs}
\begin{array}{l}
    \begin{tikzpicture}[every node/.style={scale=1},scale=0.45]
        \draw  (-0.5, 0) -- (-1.25, 0) node [left] {$\Psi  =\quad  \cdots $ } 
              (0.5, 0) -- (1.5, 0) 
              (2.5, 0) -- (3.5, 0)
              (4.5, 0) -- (5.25, 0) node [right] {$\cdots$} ;
              
        \draw (0,0) circle (0.5) node {$A$} 
              (0,0.5 ) -- (0,1.25 ) ;
              
        \draw (2,0) circle (0.5) node {$A$} 
              (2,0.5) -- (2,1.25 ) ;
              
        \draw (4,0) circle (0.5) node {$A$} 
              (4,0.5 ) -- (4,1.25 ) ;
    \end{tikzpicture}%
\end{array} ,
\end{equation}
 where $A$ is a local tensor defined for each unit cell. If a unit cell contains $M$ sites, we can further decompose $A$ as a product of $M$ local tensors defined by $A_l$ ($l=1,\dots , M)$
\begin{equation} \label{eq:gs_1001}
\begin{array}{l}
    \begin{tikzpicture}[every node/.style={scale=1},scale=0.5]
        \draw (-1.25, 0 ) -- (-0.5, 0 )  (0,0) circle (0.5) node {$A$}
              (0.5, 0 ) -- (1.25, 0 ) node [right] {\, $=$ }
              (0,0.5) -- (0, 1.25) ; 
    \end{tikzpicture}%
\end{array}
\begin{array}{l}
    \begin{tikzpicture}[every node/.style={scale=1},scale=0.55]
        \draw (-1.5, 0) -- (-0.5, 0)  (0, 0) circle (0.5) node {$A_1$} 
              (0.5, 0) -- (1.5, 0) 
              (2, 0) circle (0.5) node {$A_2$} 
              (2.5, 0) -- (3.25, 0)  (4,0) node {$\cdots$}
              (4.75, 0) -- (5.5, 0)  
              (6, 0) circle (0.5) node {$A_M$}
              (6.5, 0) -- (7.25, 0); 
              
        \draw (0, 0.5) -- (0, 1.25) 
              (2, 0.5) -- (2, 1.25)
              (6, 0.5) -- (6, 1.25);
    \end{tikzpicture}%
\end{array}
.
\end{equation}
 For the above Laughlin state, $M=3$. For the MR state at $\nu=1/2$ filling, the minimal orbital configuration to achieve a ground state of even parity is $\{...1001 1001...\}$, allowing us to use $M=4$. 
 
 However, MR is a system of pairing composite fermions in analogy to a p-wave superconductor. The neutral fermion only occurs in the system with an odd number of fermions, so we take half of the cell of pairing fermions as an unpaired ground state: $\{...01...\}$ is the unpaired configuration of $\{...1001...\}$ that contains one fermion excitation. Thus, in the study of neutral fermions, we can reduce the unit cell size from $M=$ 4 to $M=2$.

The vertical leg of $A_i$ represents the local physical degrees of freedom with quantum numbers $(K_i, C_i)$, where $C_i$ denotes the electron number and $K_i$ equals the orbital momentum $\hat{K}_i$. If the MR state in the odd parity sector with a unit cell $M=2$ is studied, we should attach a one-half quantum flux $f=1/2$ to the first local tensor in each unit cell. This additional flux modifies the momentum $\hat{K}_i$ at that site to $K_i=\hat{K}_i + f$. 

Similarly, the two horizontal legs also carry quantum numbers, denoted as $(\overline{K}_i, \overline{C}_i)$. The quantum numbers on the three legs of the same tensor must satisfy the rule:
\be\label{eq:add_rule}
( \overline{K}, \overline{C})_{i,\text{left}} + (K_i, C_i) - (\overline{K}, \overline{C})_{i,\text{right}} = 0
\ee

\begin{figure}[tp]
\includegraphics[width=0.85\columnwidth]{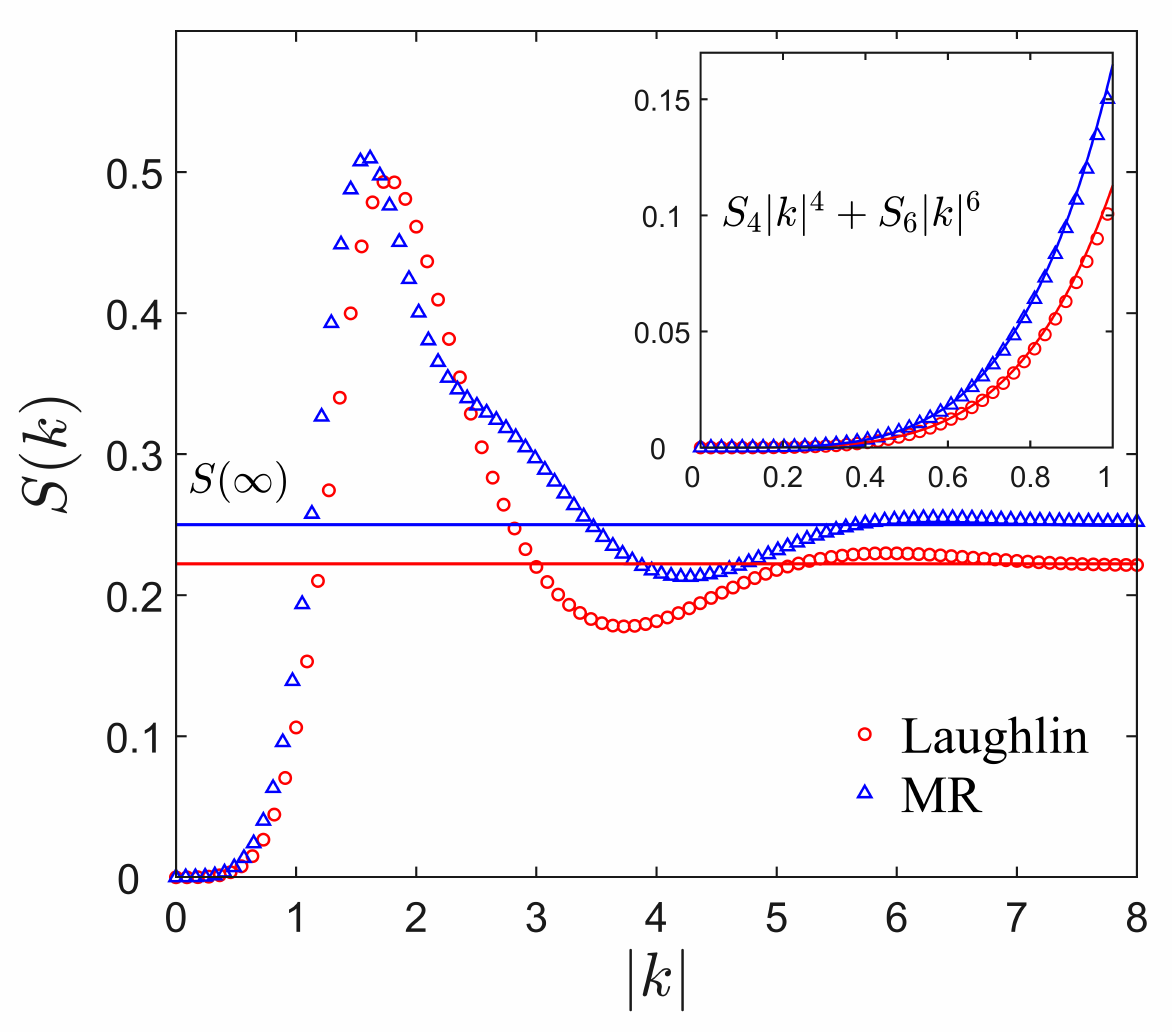}
\caption{
 Static structure factor $S(k)$ for the Laughlin and MR states along the momentum line $k=(k_x, 0)$. The solid lines are theoretical predictions $S\left(k \to \infty\right) = \nu(1 - \nu)$ with $\nu$ the filling factor \cite{Haldane2011_1}. The inset shows the structure factor in the long-wavelength limit. The solid curves are the field theory results with $S_4 = \nu s/4$ and $S_6 = \nu s [ s - (c-\nu)/ (12 \nu s) ]/8$, $s$ the guiding center spin and $c$ the central charge \cite{Kalinay2000, Can2014, Can2015, Nguyen2017, Gromov2017, Gromov2017_1, Wang2019, Kumar2024}. 
}
\label{fig:static_fac}
\end{figure}

The projected guiding center structure factor, defined by the following static density-density correlation function, characterizes the incompressibility of FQHE:
\be \label{eq:gc}
  S(k) = \frac{1}{N} \langle \delta \rho_{k }\delta \rho_{-k} \rangle ,
\ee
 where $\delta \rho_{k }=\rho_{k } - \langle \rho_{k } \rangle$ is the fluctuation of the density operator. Figure~\ref{fig:static_fac} shows the MPS results for the structure factors in the Laughlin and MR states (details are presented in SM). It indicates that both states are incompressible fluids, although their corresponding MPS representations are periodic in unit cells. In the long-wavelength limit, $S(k)$ could be expanded as 
 \be
 S(k) = S_2 |k|^2 + S_4 |k|^4 + S_6 |k|^6 + \cdots .
 \ee
In a topologically gapped system, one has $S_2=0$ and the leading contribution is from the $S_4$ term \cite{Lee1991}. 

\begin{figure*}[htp]
\includegraphics[width=2\columnwidth]{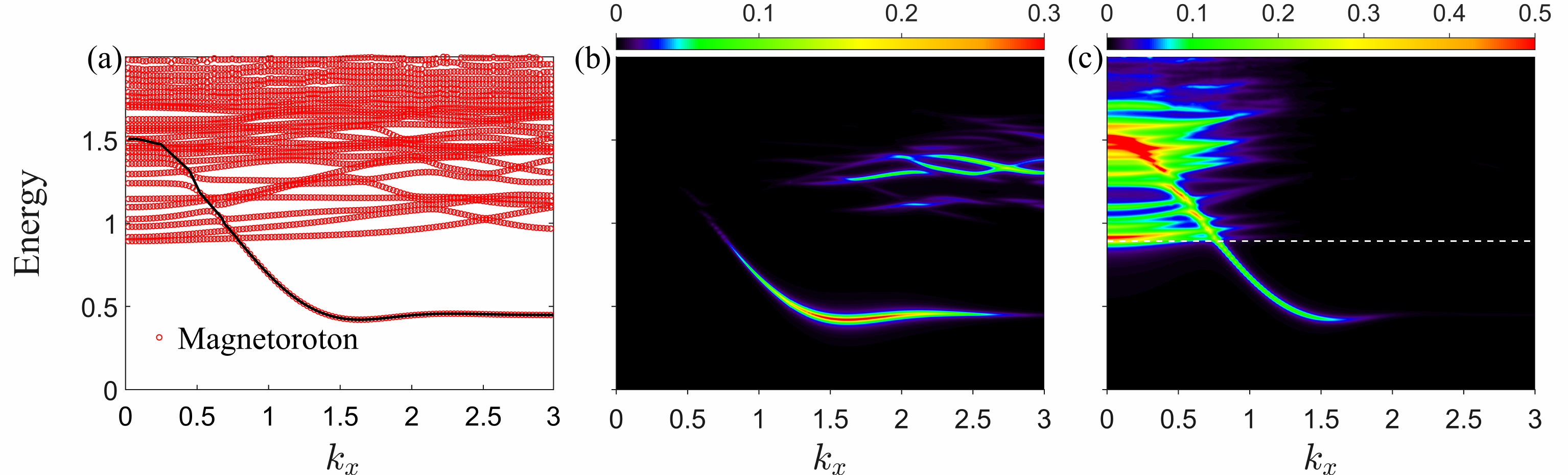}
  \caption{(a) Energy dispersions of the excited states for the $\nu =1/3$ Laughlin state on an infinite cylinder with $L_y = 18$ along the momentum line $k = (k_x, 0)$, obtained with an MPS bond dimension $D = 88$. (b) Single-particle density spectra. (c) Pairing density spectra of $O^-$. The white dashed line represents the lower bound of the continuum. The black line in (a) sketches the energy dispersion of the magnetoroton with the highest spectral weight at $(k = 0, E = 1.5)$ in (c).
}
\label{fig:laughlin}
\end{figure*}

 Under the single-mode approximation, the excited states are represented by a momentum-boosted MPS in the tangent space of $\Psi$ \cite{Ostlund1995, Rommer1997, Haegeman2011, Haegeman2012} 
\begin{equation} \label{eq:sma}
\begin{array}{l}
    \begin{tikzpicture}[every node/.style={scale=1},scale=0.45]
        \draw (-2.5,0) node [left] {$\Phi_{{k}}( B) ={\displaystyle \sum_{n}} \e^{i \tilde{k}_x nM } $} ;

        \draw (-0.5, 0) -- (-1.25, 0) node [left] {$\cdots$} 
              (0.5, 0) -- (1.5, 0) 
              (2.5, 0) -- (3.5, 0) 
              (4.5, 0) -- (5.25, 0) node [right] {$\cdots$} ; 
              
        \draw (0,0) circle (0.5) node {$ A$} 
              (0,0.5) -- (0,1.25) node [above] {$n-1$};
              
        \draw (2,0) circle (0.5) node {$ B$}
              (2,0.5) -- (2,1.25) 
              (2, 1.35) node [above] {$n$};
              
        \draw (4,0) circle (0.5) node {$ A$} ; 
        \draw (4,0.5) -- (4,1.25) node [above] {$n+1$} ;
    \end{tikzpicture}%
\end{array},
\end{equation}
where $B$ is an impurity tensor defined on the $n$th unit cell. In a unit cell of $M$ sites, $B$ is a sum of $M$ MPS, where $A_i$ in the $i$th MPS ($i=1,  \dots, M$) is substituted by a local impurity tensor $B_i$ multiplied by a site-dependent phase factor. For example, for a $M=2$ system, $B$ is defined as 
\begin{equation} \label{eq:nf}
\begin{array}{l}
    \begin{tikzpicture}[every node/.style={scale=1},scale=0.48]
        \draw (-5.25, 0) -- (-4.5, 0)
              (-4,0) circle (0.5) node {$B$} (-4, 0.5) -- (-4, 1.25)
              (-3.5,0) -- (-2.75, 0) node [right] {$= $} ; 
 
        \draw (-0.5,0) -- (-1.25,0) 
              (0,0) circle (0.5) node {$B_1$} 
              (0.5,0) -- (1.25, 0) 
              (0, 0.5) -- (0, 1.25)
              (1.75,0) circle (0.5) node {$A_2$} 
              (2.25,0) -- (3,0) 
              (1.75, 0.5) -- (1.75, 1.25) ;                

        \draw (-0.5+6.75,0) -- (-1.25+6.75,0) 
              (-1.35+7,0.1) node [left] {$+\, e^{i \tilde{k}_x}$} 
              (0+6.75,0) circle (0.5) node {$A_1$}  
              (0.5+6.75,0) -- (1.25+6.75, 0) 
              (1.75+6.75,0) circle (0.5) node {$B_2$} 
              (0+6.75,0.5) -- (0+6.75,1.25) 
              (2.25+6.75,0) -- (3+6.75,0) 
              (1.75+6.75,0.5) -- (1.75+6.75,1.25) ;                
    \end{tikzpicture}%
\end{array} .
\end{equation}

{\it Spectral functions ---}
 Upon determining all the local tensors variationally, we can use the eigenvalues and eigenfunctions of the ground and excited states to evaluate the spectral function of a physical variable $O_k$
\be\label{eq:specfunc}
I(\omega, k)=\sum_n|\langle 
\Phi_{k, n}( B)|O_{k}|\Psi( A)\rangle|^2\delta(\hbar\omega-E_n),
\ee
where $E_n$ is the $n$th eigenvalue of the excited state and $\Phi_{k, n}$ the corresponding eigenfunction. 

In the $\nu = 1/3$ Laughlin state, the lowest collective excitations are the bosonic magnetoroton modes. This mode essentially represents a quantized wave of charge density propagating through the electron system, detectable by the single-particle density spectral function of $O_k = \delta \rho_k $. As shown in Fig.~\ref{fig:laughlin}(a-b), the dispersion of this mode becomes flat in the large $k$ limit, indicating that it is a composite fermion exciton mode as predicted by Scarola {\it et. al} \cite{Scarola2000}. The spectrum shows a distinct magnetoroton minimum at  $k \approx 1.7$ with an energy $E = 0.41$, indicating a softening of this mode at a specific length scale. At small $k$, the magnetoroton mode merges into the continuum of excitations, and its spectral weight diminishes significantly, primarily due to the quartic dependence of $S(k)$ on $k$.

\begin{figure*}[htp]
\centering
\includegraphics[width=2\columnwidth]{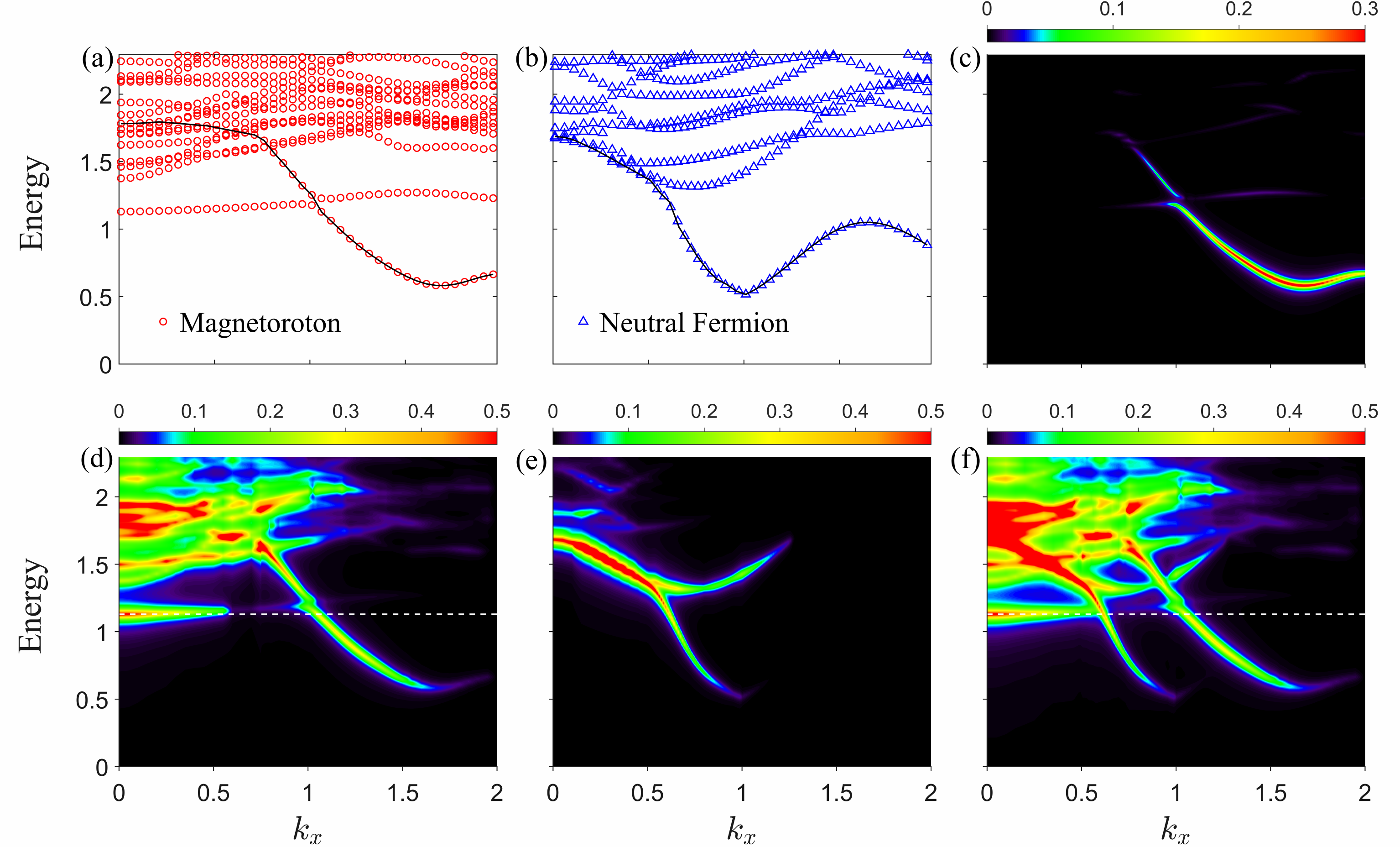}
\caption{
 Spectral functions of the $\nu = 1/2$ MR state on an infinite cylinder with $L_y=16$, obtained along the momentum points $k= (k_x, 0)$ with $D=103$. Energy dispersions of the excited states in (a) the even- and (b) odd-parity sectors. (c) Single particle density spectra. (d-e) Three-particle density spectra of $O^-$ in the even- and odd-parity sectors, respectively. (f) Combined three-particle density spectra of (d) and (e). The black curve in (a) is a guiding line that traces the high-intensity points in the small $k$ region with the magnetoroton mode shown in (c) and (d). The black curve in (b) is the energy dispersion of the lowest excited states in the odd-parity MR state. The thin white dashed lines in (d) and (f) are the lower edge of the continuum.  }
\label{fig:mr}
\end{figure*}

The single-particle density spectrum also reveals pronounced features of charge-neutral excitations, known as excitons, in the high-energy continuum, with momentum $k$ ranging from $2$ to $3$ and energy $E$ from $1.25$ to $1.4$. These features likely arise from the excitation of composite fermions to higher effective Landau levels in the composite fermion theory \cite{Das2018, Majumder2014, BoYang2014, Balram2024}. This finding aligns with earlier calculations \cite{Platzman1994} and is consistent with inelastic photon measurement results \cite{Rhone2011}.

In this Laughlin state, the intrinsic metric associated with magnetoroton couples directly with the two-particle density operators in the long wavelength limit. Thus, to reveal the magnetoroton mode in the continuum, we calculate the pair density spectrum 
Eq.~\eqref{eq:specfunc} with $O_k$ defined by 
\be\label{eq:2b_operator}
O_{k}^\pm = \sum_{k_1+k_2=k}  k_1^\pm  k_2^\pm e^{- (k_1^2 + k_2^2)/4} \delta \rho_{k_1} \delta \rho_{k_2},
\ee
where $k^\pm = k_{x} \pm i k_{y}$, corresponding to the $S=2$ and $S=-2$ representations, respectively. We also evaluate the dynamical spectra of other d-wave operators $O_{x^2-y^2} = O^{+} + O^{-}$ and $O_{xy} = O^{+} - O^{-}$. The calculation reveals that only the $S=-2$ mode contributes to the long-wavelength magnetoroton spectra in the continuum. More specifically, the $S=-2$ spectrum shows a strong signal at $k=0$ and $E\approx1.5$, as depicted in Fig.~\ref{fig:laughlin}(c), consistent with the exact diagonalization results  \cite{Liou2019}.
 
At $k=0$, the $S=-2$ spectrum also reveals a peak just above the lower edge of the continuum at $E=0.9$. This peak corresponds to a bi-roton bound state, where two rotons interact by dipole-dipole interaction to form a bound state with an energy lower than the magnetoroton mode in the continuum \cite{Platzman1996, Park2000, Ghosh2001}. However, this bound state does not possess the chirality characteristic of the geometric, aligning with recent experimental results \cite{Liang2024}. 

 Next, we turn to the excitations of the MR state. In the MR state, electrons pair up in a p-wave superconducting-like manner. In addition to the magnetoroton modes, neutral fermion modes emerge in this state. These modes can be described by Majorana fermions bound to vortex excitations in the p-wave paired state. 
 
Based on the calculations of single and three-particle density spectra, we identify two types of excitations,  as shown in Fig.~\ref{fig:mr}(a-b). The first type is the bosonic magnetoroton modes, which exhibit energy dispersions similar to those in the Laughlin states. The second type is the neutral fermion modes, which features a distinct local minimum at $k_x = 1$. Our results are consistent with previous finite-system calculations based on the bipartite composite fermions, Jack Polynomials, symmetrization constructions on multilayer systems, and supersymmetric wave functions \cite{Sreejith2011, BoYang2012_2, Repellin2015, Pu2023}.

 The single-particle density spectrum, depicted in Fig.~\ref{fig:mr}(c), shows similar results to those in the Laughlin state (see Fig.~\ref{fig:laughlin}(a)). Notably, the low-energy magnetoroton excitation exhibits a prominent peak at $k_x \approx 1.6$ and $E=0.57$. However, the spectrum becomes heavily damped upon entering the two-particle excitation continuum. Like the Laughing state, it shows no spectral weight for the magnetoroton modes in the long-wavelength limit. 

 Unlike in the Laughlin state, the magnetoroton mode does not display a sizable weight in the pair density spectra in the long-wavelength limit due to the paired nature of electrons in the MR state. This pairing modifies the geometry and metric that govern the interactions and correlations. Consequently, the magnetoroton modes do not impact the MR state similarly to the Laughlin state, especially in the pair density channel. 

 However, the magnetoroton modes respond strongly to the dynamic fluctuation of the three-particle density operator  
\begin{eqnarray}\label{eq:3b_operator}
  O^\pm_{k}&=& \sum_{k_1k_2k_3} k_1^\pm ( k_2^\pm + k_3^\pm )  e^{-(k_1^2 + k_2^2+ k_3^2)/4} \nonumber
  \\
  && \qquad\delta \rho_{k_1} \delta \rho_{k_2}\delta \rho_{k_3} \delta_{k_1+k_2+k_3, k}.
\end{eqnarray}
 Again, $O^\pm$ corresponds to the $S=\pm 2$ state. Figure~\ref{fig:mr}(d) illustrates the magnetoroton spectra acquired in the $O^-$ channel. Similar to the Laughlin state, there is a distinct magnetoroton response in the small $k$ region within the continuum. Compared to the $S=-2$ mode of the Laughlin state, the spectrum of this mode in the MR state is more broadened in the low-$k$ region, consistent with the result presented in Ref.~\cite{Liou2019}. 

 Magnetoroton modes are linked to collective density oscillations, whereas neutral fermion modes are closely associated with the topological excitations of the MR state. To explore the neutral fermion modes, we compute the MR ground state and the corresponding excited states in the odd parity sector. The analysis of response functions reveals that both the pair and three-particle density spectral functions exhibit distinct peaks of neutral fermions. Figure~\ref{fig:mr}(e) shows the three-particle density spectra in the neutral fermion channel. This is the first time the spectral function of neutral fermions in the small $k$ limit is obtained. As illustrated by Fig.~\ref{fig:mr}(f), the magnetoroton and neutral fermion modes tend to merge at $k=0$, hinting at an underlying emergent supersymmetry. Furthermore, the energy of the magnetoroton mode is higher than that of the neutral fermion mode in the continuum, consistent with the published result for the two modes obtained by parametrizing the superspace \cite{Pu2023}. 

\par \textit{Conclusion and discussion---}
In summary, we investigate the neutral excitations of the $\nu =1/3$ Laughlin state and the $\nu = 1/2$ MR state under the single-mode approximation of MPS on an infinite cylinder. Our analysis of the Laughlin state pinpoints the magnetoroton minimum and elucidates the long-wavelength magnetoroton mode within the continuum through the pair density spectral function. From the long-wavelength spectra, we confirm that the magnetoroton mode corresponds to the $S=-2$ geometric excitation.

 For the MR state, we have developed the MPS representations for both ground and excited states across even and odd parity sectors. We probe the bosonic magnetoroton and neutral fermion modes in the MR state in the small $k$ region, inaccessible via exact diagonalization due to the finite-size effects. Our results confirm that these modes are consistent with the previously proposed emergent supersymmetry. It is worth exploring whether incorporating supersymmetry into effective field theory might pair these two modes as superpartners \cite{Gromov2017}. The dynamic spectra we obtain will also contribute to constructing corresponding massive wave equations for these fields.

Our calculation is based on the parent Hamiltonians, which account only for short-range interactions, not the long-range Coulomb interaction. Nevertheless, we believe that our findings still offer significant qualitative insights into the physics underlying FQHE. Our results suggest that the neutral fermion mode might be explored experimentally through multi-photon experiments, providing a new avenue for empirical verification.

\textit{Acknowledgments---} %
We have implemented the code for the excited spectrum based on ITensor \cite{Fishman2022}. We thank Songyang Pu, Zlatko Papi\'{c}, Kun Yang, Ying Hai Wu, Tong Liu, Lingjie Du, Ajit C. Balram and J. K. Jain for helpful discussions. We thank J. Haegeman for the instructions on the KrylovKit package. This work is supported by the NSFC grant No. 12488201, and by the China Postdoctoral Science Foundation grant No. 2023M743742.

\bibliography{main}

\newpage

\clearpage

\addtolength{\oddsidemargin}{-0.75in}
\addtolength{\evensidemargin}{-0.75in}
\addtolength{\topmargin}{-0.725in}

\newcommand{\addpage}[1] {
 \begin{figure*}
   \includegraphics[width=8.5in,page=#1]{Appendix.pdf}
 \end{figure*}
}
\addpage{1}
\addpage{2}
\addpage{3}
\addpage{4}
\addpage{5}

\end{document}